# Tag Clouds for Object-Oriented Source Code Visualization

Ra'Fat Al-Msie'deen
Department of Computer Information Systems, Faculty of IT
Mutah University, Karak, Jordan
rafatalmsiedeen@mutah.edu.jo

*Abstract*—Software visualization helps software engineers to understand and manage the size and complexity of the object-oriented source code. The tag cloud is a simple and popular visualization technique. The main idea of the tag cloud is to represent tags according to their frequency in an alphabetical order where the most important tags are highlighted via a suitable font size. This paper proposes an original approach to visualize software code using a tag cloud. The approach exploits all software identifier names to visualize software code as a tag cloud. Experiments were conducted on several case studies. To validate the approach, it is applied on NanoXML and ArgoUML. The results of this evaluation validate the relevance and the performance of the proposed approach as all tag names and their frequencies were correctly identified. The proposed tag cloud visualization technique is a helpful addition to the software visualization toolkit. The extracted tag cloud supports software engineers as they filter and browse data.

## I. INTRODUCTION

Object-oriented (OO) source code visualization helps software engineers understand and manage the complexity and size of software code [1]. This paper proposes an original approach called Iconic to visualize OO software as a tag cloud [2] (the name of approach is inspired by icons, iconic is the feature of an icon). A tag cloud is a visual representation of textual content that uses color and size to point out word frequency [3]. The tag is generally a single word. The importance of each tag is displayed with font size or colour [4]. Tag clouds can be categorized into two categories: static tag clouds and dynamic tag clouds [5]. Most existing approaches are designed to extract tags from web pages and text documents [3, 5-7]. The current studies that identify tag clouds from software code exploit only software classes and methods [8-11]. Moreover, these approaches add class and method names to the tag cloud without any pre-processing (i.e. as written in the software code). In the literature, there is no approach to identify tag clouds by using all software identifiers (i.e. packages, classes, attributes and methods).

In this paper, tag cloud shows the most common tags across software identifiers. In the cloud some tags may appear more important than others, where the tag frequency determines the tag font size. The use of color is arbitrarily, it is used just for aesthetic purposes. Tags are sorted alphabetically or according to their frequency (Figure 2). Iconic accepts the source code of software systems as input. Then, based on static code analysis [12], Iconic extracts all software identifier names. Then, it splits the identifier names into their constituent words. Then, it acquires the words' roots. After that, it assigns weights to each tag based on its frequency across software code and stores the tags in a standard order. Finally, Iconic builds the tag cloud as output.

## II. RELATED WORK AND COMPARISON WITH ICONIC

Authors in [9] used tag cloud to visualize software classes. The extracted tag clouds exposed the most common tags used in software class names. Iconic visualizes software package, class, attribute and method names as a tag cloud. The identified tag cloud has exposed the most frequently-used tags across software identifier names. Authors in [10] proposed an approach to visualize software methods via tag cloud. The tag cloud visualizes names of methods, parameters and local variables. In a tag cloud, if a tag name is selected, then the related source code elements in the graph visualization will be highlighted. In Iconic, there is no graph visualization of source code elements. Iconic does not identify the link between tag in the cloud and the source code elements in the graph visualization. Iconic visualizes software identifiers as a tag cloud. Authors in [8, 11] used the tag cloud to visualize software methods. Their tool allows the user to explore the tag cloud using different layouts. In a tag cloud if a tag is selected, then the related tags in the cloud will be highlighted. In their work, the tag cloud is customizable and the tool allows tag layouts to change. Iconic uses a tag cloud to visualize all software identifiers. Iconic allows the software developer to explore the tag cloud using a typewriter layout with tag names in alphabetical order. Author in [13] presents an automatic approach to extract software code labels. The approach splits the name of an identifier into a set of keywords. Then, it returns each keyword to its stem to generate the code labels. The approach creates labels with the same font size and colour. There is no indication about label importance in the extracted code labels. In Iconic, the tag frequency determines the tag font size in the tag cloud.

There are other approaches to visualize software code by using different techniques. In a previous work [1], this author has developed a tool named Vsound to visualize the software source code and its main dependencies. Vsound depends on software identifiers to visualize, understand and document the software code. Vsound aims for a graphical representation of software code as a graphic-based document. Authors in [14] suggested a feature naming technique by using the VariClouds approach. VariClouds is a method that uses word cloud visualization in order to assign names to identified blocks (i.e. features) based on the most frequent words in those blocks. VariClouds is developed for helping software engineers in feature identification and naming. Authors in [15] offered a feature naming method as part of their automatic feature model extraction technique (called REVPLINE approach). They assigned names for features based on the most frequent tokens of the identified blocks. The most frequent tokens within the block represent the most frequent tags within a tag cloud. The word cloud can be used in [15, 16] in order to identify the name of the features extracted from the software code as atomic blocks based on the most frequent words in those blocks. In a software engineering domain, there are limited existing software engineering tools which use a tag cloud visualization technique. Eclipse plugin Sourcecloud [17] creates a tag cloud visualization of the text within a package, class or project with font size weighted by tag frequency and colors assigned arbitrarily. On the other hand, Iconic uses all software identifiers to generate tag cloud, not the whole software document. A tag cloud is often used to show an overview of the contents of textual documents. Wordle [18] made it easy to generate tag clouds from a specific text. Wordle is a tag cloud generator for any text. Wordle presents tags in the cloud without any pre-processing; as it appears in the text, while Iconic uses WordNet [19] to do some simple pre-processing on the software identifiers such as stemming and removal of stop words. Most existing approaches are designed to extract tag clouds from web pages and textual documents. In software systems, there are limited existing approaches which use a tag cloud to visualize software code. The current approaches use tag clouds to visualize software classes or methods. The concise overview of the existing approaches shows the need to propose an approach to extract tag cloud from software code using all software identifier names.

### III. APPROACH OVERVIEW

This section presents the main ideas used in Iconic. It also gives an overview of the tag cloud process. Finally, it shortly describes the example that illustrates the tag cloud process. The main objective of Iconic is to visualize software identifiers as a tag cloud. The tag cloud displays the most frequently-used words across software identifiers. Tag cloud builds an alternative representation of the software identifiers at a higher level of abstraction. Figure 1 presents the tag clouds process and a sample execution of the Iconic approach (i.e. DrawingShapes). This process takes the software source code as its input. Its first step aims to identify all software identifiers. Then, Iconic splits the identifier names into their constituent words. In the next step, Iconic turns the identifier words into their word stems or roots. Then, Iconic assigns a weight for each tag based on its appearance frequency. After that, Iconic stores all tags in a standard (e.g. alphabetical) order. Finally, Iconic generates the tag cloud.

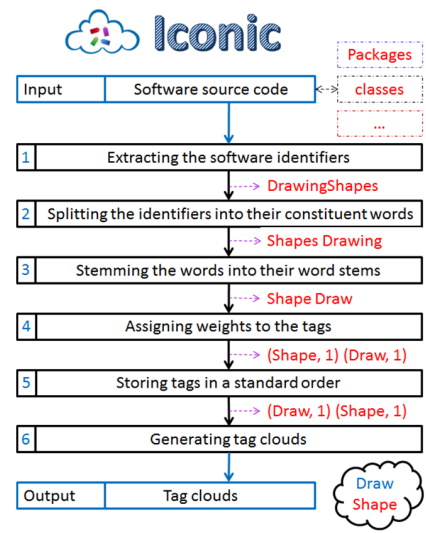

Fig. 1.    The tag cloud process, and a running example.

As an illustrative example, Iconic approach considers the drawing shapes software system [12, 13]. This software product allows the software developer to draw three types of shapes which are: lines, rectangles and ovals. In addition, drawing shapes software lets the user to select the shape color.

### IV. TAG CLOUD PROCESS STEP BY STEP

Iconic identifies the software tag cloud in six steps which are detailed below.

#### A. Extracting Software Identifiers

Iconic takes the software code as input. Then, Iconic parser generates an identifier file as output. Identifier file contains the main OO elements (i.e. package, class, attribute and method). Iconic uses the eclipse Java development tools and the eclipse abstract syntax tree to access, modify and read the elements of the software [20]. The abstract syntax tree is widely used in numerous areas of software engineering as a representation of source code.

#### B. Splitting the Identifiers into their Constituent Words

Iconic splits the software identifier names into a set of words. Iconic uses the camel-case splitting algorithm [15], which splits words based on capital letters, underscores and numbers. Each identifier name is split into words based on the camel-case syntax. For example: DrawingShapes is split into drawing and shapes. Camel-case method is a simple and broadly used method for identifier splitting algorithms [21] and the rules of splitting are largely based on camel-case convention.

#### C. Stemming Words into their Word Stems

Stemming is a method of stripping affixes from words to form the word stem or base (e.g. performed to perform). In Iconic approach, the stemming (e.g. removing word endings) was achieved via WordNet [19]. Iconic uses WordNet [22] dictionary to replace English words with their stems. In Iconic, stemming is a way of converting an identifier name (word) to its stem (tag). For instance, the words writing, wrote and written all have the same base/stem which is write.

*D. Assigning Weights to Tags*

In Iconic approach, the tag weight gives an indication about the tag frequency (or tag importance) across software identifiers. In this step, tag weight is assigned to each tag, based on its frequency of appearance in software identifiers. For example, in drawing shapes software the draw tag occurred ten times across software identifiers, so the given weight of this tag is ten (Figure 4).

*E. Storing Tags in a Standard Order*

In Iconic, the tags within a tag cloud are arranged from left to right, and top to bottom (i.e. typewriter style). Iconic presents tags in a tag cloud in alphabetical order (Figure 2). The software engineers seem more able to simply find tags in alphabetically ordered clouds [11]. Authors in [8] indicated that alphabetical ordering is particularly effective for tasks involving searching for a specific tag or confirming its absence from the cloud.

*F. Generating Tag Clouds*

Iconic approach generates several clouds based on the software identifiers. Iconic clouds cover all granularity levels of the software code. Figure 2 shows the tag cloud extracted from drawing shapes code. This tag cloud contains all software identifier tags (i.e. package, class, attribute and method). The tag cloud in Figure 2 shows that the most frequently-used words (tags) in software identifier names are draw and shape. The most frequent tags are shown in larger fonts.

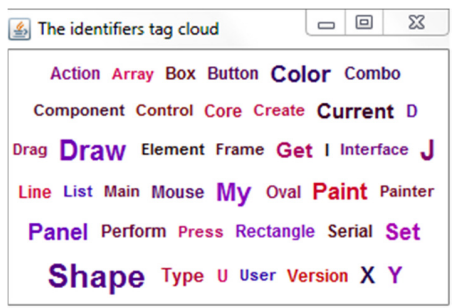

Fig. 2.    Tag cloud generated from drawing shapes software.

Tag cloud helps software engineers to get the most common and rare tags. Using Iconic approach five tag clouds are generated: these are package, class, attribute, method, and all software identifiers. Different software tag clouds are shown in Figure 3. Filtering the unwanted tags from tag clouds is a critical issue. Many filters can be added to the tag cloud [23]. The cloud filters are very important for filtering the unwanted tags such as short tags. On the other hand, cloud filters can be added to present precise information about tags such as the number of tag frequency. Filtering facilities are essential and very important to deal with enormous tag clouds. Iconic uses two filters: short-tag and tag-frequency filters. The short-tag filter aims to filter out the tags which have less than four characters. While, the tag-frequency filter can be used as an indicator for the tag frequency across software identifiers.

These two filters are very useful for software developers. For example, short-tag filter is very important when there are too many tag names to display in the cloud. By using this filter some of the smaller names will not be displayed in the cloud, while the tag-frequency filter determines the frequency of the tag name across software identifier names as a precise number between square brackets. Figure 4 presents the generated tag cloud from all identifiers of drawing shapes software after applying these two filters (Figure 2 shows the generated tag cloud from the same software without filters).

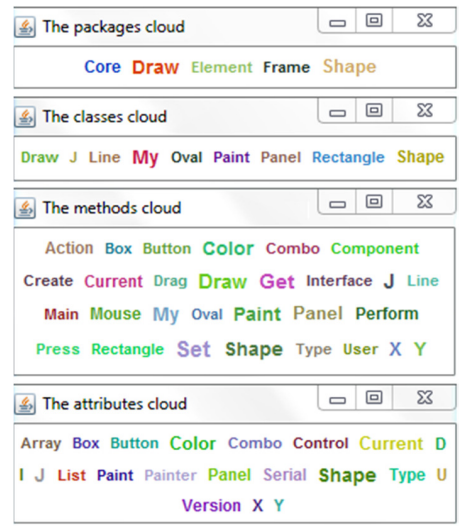

Fig. 3.    Different tag clouds extracted from Drawing Shapes software.

In the current approach, the tags are produced stripped of information as they are simply word stems of split identifiers. The whole identifiers may carry some information (e.g. DrawingShapes) but the split and stemmed tags do not carry the same information (e.g. Draw and Shape). Figure 5 shows the identifiers tag cloud of the drawing shapes program without splitting and stemming the identifier names. White, black and red colours are used as a design choice of the cloud. The tag labels are shown in black, while the tag frequencies are presented in red color between square brackets.

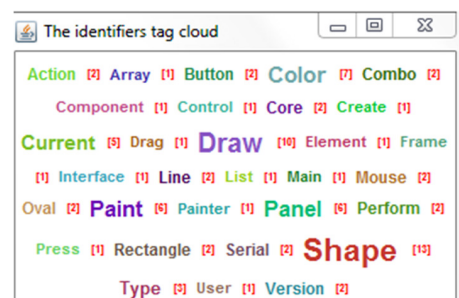

Fig. 4.    A tag cloud generated by using Iconic filters.

V.  EXPERIMENTATION

This section presents the experiments conducted in this study to show its soundness, and presents the ArgoUML and NanoXML case studies. It also summarizes the obtained results for the case studies, presenting the threats to the validity of Iconic approach. ArgoUML [24] is a Java-based software. ArgoUML is used for designing software systems in UML. NanoXML software [25] is a Java program for parsing XML files. Iconic implementation [26] converts software code into a tag cloud. ArgoUML software represents a large system [27]. Iconic performed an evaluation of the execution time (in ms) of its algorithms. Table I presents the execution time for each case study. In addition, software size and the number of software identifiers are presented. ArgoUML software is considered as a large base code (i.e. 120,348 lines of code). The execution time of Iconic on this case study is relatively little (approximately 45s). The different size and complexity levels show the capability of Iconic to deal with such software systems.

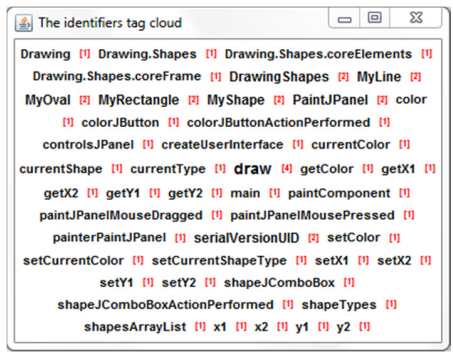

Fig. 5.   Tag cloud visualization created by Iconic.

TABLE I.   INFORMATION ABOUT TAG CLOUDS EXTRACTED FROM CASE STUDIES.

| Case study | ArgoUML | NanoXML |
|---|---|---|
| # of packages | 103 | 3 |
| # of classes | 1745 | 24 |
| # of attributes | 3649 | 63 |
| # of methods | 10319 | 318 |
| # of identifiers | 15816 | 408 |
| # of tags | 1511 | 135 |
| Execution time | 44836 | 1197 |

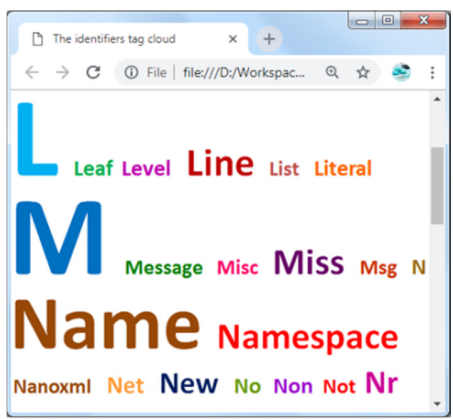

Fig. 6.   Tag cloud extracted from NanoXML software.

Figure 6 shows the generated tag clouds from NanoXML software. The most frequent tag names in ArgoUML are Get and Action, while in NanoXML are Get, and Attribute. Figure 7 shows the tag cloud generated from the ArgoUML after applying the short-tag and tag-frequency filter.

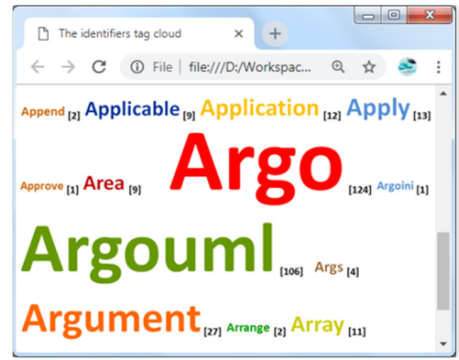

Fig. 7.   Tag cloud extracted from ArgoUML software.

The extracted tag clouds from ArgoUML and NanoXML show that the developer can easily find tags in alphabetically ordered clouds [8]. Software developers look able to more simply find tag names in alphabetically ordered clouds [28]. The evaluation of tag cloud is a vital, but challenging, aspect of visualization [8]. Despite the abundance of visualization techniques proposed in the software engineering field, there remains a dearth of extensive use of the tag cloud technique. To evaluate the proposed filters, a simple case study with five Java developers as participants was performed. Upon starting the evaluation, each participant was asked to see the tag cloud of NanoXML software without filters (Figure 6) and with filters (Figure 8). Then, each participant was asked to answer two questions with agree or disagree. Table II displays the study design in detail.

TABLE II.   THE DESIGN OF THE USER STUDY. ALL PARTICIPANTS SEE DIFFERENT CLOUDS FOR THE SAME SOFTWARE.

| Kind of cloud | | Question # | Users' ratings for each question S1: Agree S2: Disagree | | | | |
|---|---|---|---|---|---|---|---|
| Without filters | With filters | | P1 | P2 | P3 | P4 | P5 |
| × | | Q1 | S2 | S2 | S2 | S2 | S2 |
| × | | Q2 | S1 | S2 | S1 | S2 | S2 |
| | × | Q1 | S1 | S1 | S1 | S1 | S1 |
| | × | Q2 | S1 | S1 | S1 | S1 | S1 |
| Questions asked: | | | | | | | |
| Q1 | The cloud is missing important tag names. | | | | | | |
| Q2 | The cloud contains information that helps me understand the importance of tag names. | | | | | | |

The cloud without filters contains all tag names while, the tag cloud with filters does not contain all the tag names, because the smaller tag names are omitted. Also, the tag cloud with filters shows the importance of tag names using tag-frequency in addition to tag font size. Figure 8 shows the generated tag cloud from NanoXML software using Iconic filters. The effectiveness of Iconic approach is measured by their precision, recall and F-Measure [20]. For a given tag name within the cloud, precision is the ratio of correctly retrieved tag frequencies to the total number of retrieved tag frequencies, while recall is the ratio of correctly retrieved tag frequencies to the total number of relevant tag frequencies. F-Measure is the harmonic mean between precision and recall [29]. F-Measure gives a high value in cases where both precision and recall are high [30]. All metrics have values between zero and one. Table III summarizes the obtained results for some samples.

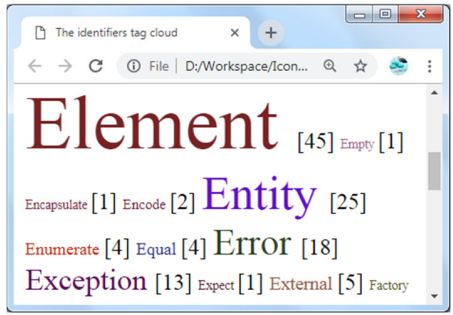

Fig. 8.   A tag cloud generated from NanoXML software.

TABLE III.   TAGS MINED FROM NANOXML AND ARGOUML

| Case study | Tag name | * | ** | Evaluation Metrics | | |
|---|---|---|---|---|---|---|
| | | | | Precision | Recall | F-Measure |
| ArgoUML | Apply | 13 | 13 | 1 | 1 | 1 |
| | Area | 9 | 9 | 1 | 1 | 1 |
| | Array | 11 | 11 | 1 | 1 | 1 |
| NanoXML | Exception | 13 | 13 | 1 | 1 | 1 |
| | Element | 45 | 45 | 1 | 1 | 1 |
| | Entity | 25 | 25 | 1 | 1 | 1 |

* Tag frequency within the cloud ( Figure 7 and 8)
**The number of identifiers that contain this tag

Results show that the precision value is one for all tag names and their frequencies. If precision equals one, all retrieved tag frequencies are relevant. This means that all generated tag names and their frequencies are correct. Moreover, tag cloud does not miss any tag names. This accuracy in the result is due to the pre-processing steps, where the identifier names are split into multiple words based on the camel-case splitting algorithm. Then, word stemming is applied to find the root of each word. Considering the recall metric, recall value equals one for all tag names. If recall equals one, all relevant tag frequencies are retrieved. This means that all frequencies that measure the tag importance are counted. Considering the F-Measure metric, the F-Measure value equals one for all tag names. If F-Measure equals to one, all relevant tag frequencies are retrieved, and only relevant tag frequencies are retrieved. This means that all frequencies that display the importance of tag names via font size are extracted. The result shows the efficiency of Iconic approach.

The threat to the validity of Iconic is that the current prototype considers only Java software. Furthermore, when a software developer uses mix words to name software identifiers (e.g. SeTSettingS) the camel-case splitting algorithm can't handle it. The WordNet may not be reliable in all cases to find the word stem. Currently, the tag clouds are missing some filters for example, they do not filter tag names that are too long (e.g. filter the identifier names to only 5 characters). In addition, it would be much faster to be able to have a filtering search that the developer can type the tag to find it.

## VI.  CONCLUSION AND PERSPECTIVES

This paper proposed an original approach to visualize all software identifiers as a tag cloud. Iconic was implemented on numerous case studies. Iconic has been applied on NanoXML and ArgoUML software. Results showed that all tag names and their frequencies were correctly identified. The extracted tag clouds have shown the most common and rare tags. Tags are sorted alphabetically or according to their frequency. The most frequent tags are highlighted via appropriate font size and color. Tags within the cloud are filtered according to their length or frequency. For future work, Iconic is planned to support the current tag cloud with a set of user tasks [23], including tag searching, browsing, zooming and filtering. It is also planned to use other layouts for tag cloud such as spiral layout [31]. In addition, it is planned to build a tag cloud using software identifiers and code comments [32] and to identify the link between tag names and software identifiers. Thus, in a tag cloud if a tag name is selected, then the related identifiers will be highlighted. Therefore, the software developers could click a tag name and see where it is used in the source code. Finally, the use of JavaDocs in Iconic [33] to build the tag cloud is another future work aim.

AUTHOR PROFILE

**Ra'Fat Al-Msie'Deen** is an Assistant Professor at Mutah University since 2014. He received his PhD in Software Engineering from the University of Montpellier 2, Montpellier – France, in 2014. He received his MSc in Information Technology from the University Utara Malaysia, Kedah – Malaysia, in 2009. He got his BSc in Computer Science from Al-Hussein Bin Talal University, Ma'an – Jordan, in 2007. His research interests include software engineering, software product line engineering, and formal concept analysis.